\documentclass[twocolumn,showpacs,preprintnumbers,amsmath,amssymb,aps,pra,tightenline]{revtex4}

\usepackage{graphicx}
\usepackage{dcolumn}
\usepackage{bm}

\usepackage{color}
\begin{document}

\title{Non-equilibrium thermodynamics approach to open quantum systems}

\author{Vitalii Semin$^1$ and Francesco Petruccione$^{1,2}$ }

\affiliation{{$^1$Quantum Research Group, School of Chemistry and Physics,
 University of KwaZulu-Natal, Durban, 4001, South Africa},
 {$^2$National Institute for Theoretical Physics (NITheP, KZN), KwaZulu-Natal, South Africa}}

\date{\today}

\begin{abstract}
Open quantum systems are studied from the thermodynamical point of view unifying the principle of maximum informational entropy and the hypothesis of relaxation times hierarchy. The result of the unification is a non-Markovian and local in time master equation that provides a direct connection of dynamical and thermodynamical properties of open quantum systems. The power of the approach is illustrated with the application to the damped harmonic oscillator and the damped driven two-level system resulting in  analytical expressions for the non-Markovian and non-equilibrium entropy and inverse temperature.
\end{abstract}


\maketitle

\section{Introduction} 
Open quantum systems attract attention in both theoretical \cite{PURI,Repke2,GARDINER} and experimental \cite{BAR, Diehl} research. The theoretical investigation of open quantum systems uses a great variety of different tools \cite{toqs}. Typically, these methods allow the determination of a reduced density matrix \cite{toqs,JENS,JPAZ} or a quantum state vector \cite{QDOS,KOSS} of an open quantum system. Usually, the reduced density operator carries also unnecessary information about different correlations in the open quantum system that are inaccessible to experimental observation. To study only the relevant information about an open quantum system it is appropriate to use methods of non-equilibrium thermodynamics.

As for the theoretical study of open quantum systems, several approaches to non-equilibrium thermodynamics have been proposed. These include methods based on projection operators by Robertson \cite{ROBERT}, Mori \cite{MORI}, Kawasaki and Gunton \cite{KAWASAKI}, Zubarev's non-equilibrium statistical operator \cite{ZUB61,ZUBKAL}, or more recently the GENERIC formalism  by \"{O}ttinger and Grmela \cite{OTIN1}. At first sight these methods seem to differ from each other and  often they are different  from canonical methods  used in the theory of open quantum systems \cite{toqs}. Nevertheless, all the methods mentioned above are based on similar ideas  and can be used for the description of open quantum systems.
 
In this article we apply concepts of non-equilibrium thermodynamics to open quantum systems and derive an alternative local in time non-Markovian master equation. After that, we illustrate the formalism with the description of the dynamics and thermodynamics of two open quantum systems, namely, the damped harmonic oscillator and the damped driven two-level system. 

The article is organised as follows. In section II we describe the main ideas of non-equilibrium thermodynamics and derive a new type of local in time master equation. The short section III deals with thermodynamical relations, which follows from the considered formalism. In section IV we apply our formalism to a general open system. In sections V and VI we study two concrete open systems, namely, the damped harmonical oscillator and the driven two level system. Finally, in section VII we conclude. Some additional details are discussed in several appendices.
 
\section{Thermodynamics of non-equilibrium systems} 
Non-equilibrium thermodynamics is based on two general assumptions \cite{Repke}: (i) The non-equilibrium macroscopic state is specified by a set of observables which are the average values $\langle P_m \rangle^t$ of some relevant dynamical variables $P_m$, where the superscript $t$ indicates time-dependence. These variables give the reduced description of the system on the chosen time scale. The existence of different time scales is due to the ``hierarchy'' of basic relaxation times in macroscopic systems; (ii)  In order to obtain a closed system of generalized transport equations for the observables, we have to construct a special solution of the Liouville equation, which is a functional of these observables.

The hypothesis of the ``hierarchy'' of basic relaxation times implies that a relaxation of a system  goes sequentially through several stages. The closer the system comes to equilibrium the fewer relevant variables are necessary. In many cases the equilibrium state is defined only by one parameter, namely, the free energy or the thermodynamically conjugate entropy. Further away from equilibrium, the number of relevant variables increases. Once a set of relevant variables is chosen, one can move to the second assumption above. 

Here we construct the relevant distribution from the principle of maximum information entropy $S=-\mathrm{Tr}\rho\ln\rho,$ where $\rho$ is the density operator of the system. This idea lies at the basis of many methods of non-equilibrium thermodynamics \cite{ROBERT,KAWASAKI,ZUB61,ZUBKAL,MORI}, but, traditionally, is ignored in the theory of open quantum systems.

The relevant distribution, which maximizes the information entropy, has the following form
\begin{equation}\label{reldist}
\rho_{\mathrm{rel}}(t)=\exp\left\lbrace -\Phi(t)-\sum_m F_m(t)P_m \right\rbrace,
\end{equation} 
where the Massieu-Planck function $\Phi(t)$, is determined from the normalization condition for the relevant distribution and has the form
 \begin{equation}\label{MP}
\Phi(t)=\ln \mathrm{Tr}\exp\left\lbrace \sum_m F_m(t)P_m \right\rbrace.
 \end{equation}
The Lagrange multipliers $F_m(t)$ are determined from the self-consistency conditions
\begin{equation}\label{scond}
\langle P_m \rangle^t=\langle P_m \rangle^t_\mathrm{rel}\equiv \mathrm{Tr}(P_m\rho_\mathrm{rel}(t)).
\end{equation}

To derive the dynamical equation for $\rho_\mathrm{rel}$ it is effective to use some kind of projection operator $\mathcal{P}$, which  satisfies the condition $\mathcal{P}\rho=\rho_\mathrm{rel},$ where $\rho$ is an arbitrary solution of the Liouville equation. One of the most general types of projection operators, %which action on the arbitrary density operator gives \eqref{reldist}, 
was introduced by Kawasaki and Gunton \cite{KAWASAKI}. The time-dependent Kawasaki-Gunton projection operator $\mathcal{P}(t)$ acts on an arbitrary operator $A$ as 
\begin{equation}\label{kawas}
\mathcal{P}\!(t)A\!\!=\!\!\rho_{\mathrm{rel}}\!(t)\mathrm{Tr}A\!+\!\!\!\sum_m\!\! \left\lbrace\mathrm{Tr}\!(AP_m)\!\!-\!\!(\mathrm{Tr}\!A)\!\langle\! P_m \rangle^t\!\right\rbrace\!\!\frac{\partial\rho_\mathrm{rel}(t)}{\partial \langle P_m \rangle^t}.
\end{equation}
Notice, that the Kawasaki-Gunton projection operator generalizes other types of time-dependent projection operators \cite{Repke}. For example, the Robertson \cite{ROBERT} and Mori \cite{MORI}  projectors can be easily obtained as a partial case of \eqref{kawas}. Moreover, the Kawasaki-Gunton projection operator technique can be directly connected to Zubarev's non-equilibrium statistical operator \cite{ZUB61,ZUBKAL}.

Once the projection operator is  chosen  one can apply a standard procedure to derive a master equation \cite{toqs,SM}. 
The resulting master equation  has a structure, which is analogue to the famous Nakajima-Zwanzig master equation \cite{NAKAJIMA,ZWANZIG}, but is now more complicated due to the explicit time-dependence of the projection operators. At this stage, a variety of different approaches can be used, that where developed for the simplification of the Nakajima-Zwanzig master equation. With the help of the so-called, time-convolutionless projection operator technique (TCL) \cite{toqs} the integro-differential Nakajima-Zwanzig  equation can be cast in the form of a differential equation due to the inversion of the evolution of the total system and manipulation of the integral term in the master equation. Application of the TCL technique to the Kawasaki-Gunton master equation leads to a master equation local in time of the following form
\begin{eqnarray}\label{local}
\mathcal{P}(t)\frac{\partial}{\partial t}\rho(t)=\mathcal{K}(t)\mathcal{P}(t)\rho(t)+\mathcal{I}(t)\mathcal{Q}(t)\rho(t_0),
\end{eqnarray}
where $\mathcal{K}(t)\!\!\!\!\!\!\!\!\!=\!\!\!\!\!\!\!\!\!\mathcal{P}(t)L(t)[1-\Sigma(t)]^{-1}\mathcal{P}(t),$ $\mathcal{I}(t)\!\!\!\!\!\!\!\!\!=\!\!\!\!\!\!\mathcal{P}(t)L(t)[1-\Sigma(t)]^{-1}\mathcal{G}(t,t_0)\mathcal{Q}(t_0),$ $\Sigma(t)=\int_{t_0}^t\mathcal{G}(t,s)\times$ $\mathcal{Q}(s)L(s)\mathcal{P}(s)G(t,s)d s.\!\!\!\!$ The superoperators \!\!\!\!\!\!\!\!\!\!\!\! $\!\!\!\!\mathcal{G}(t,t')\!\!\!\!=\!\!\!\!$ $\mathcal{T_-}\exp\left\lbrace\int\limits_{t'}^{t}\mathcal{Q}(\tau)L(\tau)d\tau\right\rbrace $ and $G(t,s)=\mathcal{T_+}\exp\left\lbrace\int\limits_{t'}^{t}-L(\tau)d\tau\right\rbrace$ are the propagator and the inverse propagator, while  the symbol $\mathcal{T_\mp}$ denotes the chronological (antichronological) ordering; $L(t)$ is the Liouville superoperator. 

Notice, that formally Eq.~\eqref{local}  has  the same structure of a traditional TCL master equation \cite{toqs}, except for the explicit time-dependence of the projectors. At the same time Eq.~\eqref{local} is a new type of a master equation, which has a few significant differences from the traditional one. Some of them will be discussed below.

The master equation \eqref{local} relates to the known Kawasaki-Gunton master equation \cite{KAWASAKI} like the traditional Nakajima-Zwanzig master equation \cite{NAKAJIMA,ZWANZIG} relates to the TCL master equation \cite{toqs}. Namely, (i) they have the same ranges of applicability; (ii) the differential TCL master equation is simpler to study than the corresponding integro-differential Nakajima-Zwanzig master equation; and an additional interesting fact \cite{MY1,MY2,PETR} (iii) the TCL master equation describes the exact dynamics of the systems more accurately than the corresponding Nakajima-Zwanzig master equation of the same order.For these reason we consider the equation \eqref{local} to be an interesting new type of master equations.

\section{Entropy and thermodynamics relations} 
An amazing feature of the above theory is the explicit connection between the dynamical variables and the thermodynamic parameters that resemble the relations in equilibrium thermodynamics. Let us define the entropy corresponding to the statistical operator \eqref{reldist}. From the self-consistency conditions \eqref{scond} we have
\begin{equation}\label{entropy}
S=-\mathrm{Tr}\rho\ln\rho=\Phi(t)+\sum_m F_m(t)\langle P_m\rangle^t.
\end{equation}
Using Eqs.~\eqref{entropy} and \eqref{MP} one easily finds the following relations
\begin{equation}\label{TR}
F_m(t)=\frac{\partial S}{\partial\langle P_m\rangle^t}\,\,\, \mathrm{and}\,\,\, \langle P_m\rangle^t=-\frac{\partial \Phi}{\partial F_m }.
\end{equation}
Note, that the index $m$ in the above equation may be continuous, in this case partial derivations must be replaced by functional derivatives and sums must be replaced by integrals. 

\section{Application to open systems}
Now, we turn to the construction of the relevant distributions \eqref{reldist} for open quantum systems.  Traditionally, an open system is supposed to consist of two weakly interacting parts \cite{toqs}. The weak interaction means that the correlation between the parts does not necessarily have to be taken into account and the density operators of the total system may be chosen as a direct product of density operators, corresponding to its subsystems. One part, called a thermostat or a bath, is a large system with a huge number of degrees of freedom. This subsystem is usually assumed to be in some equilibrium state. The most commonly used  thermal equilibrium state is characterized by the free energy only  and the density operator for such state is written as $\rho=\exp[-\beta H]/\mathrm{Tr}\exp[-\beta H],$ where $\beta$ is the inverse equilibrium temperature and $H$ is the Hamiltonian of the thermostat.
The second part of the total system, namely the open system itself, is generally assumed to be in some non-equilibrium state. Unfortunately, there is no universal criterion for the choice of the relevant variables for this case. For a low-dimensional system one can choose all possible dynamical variables as relevant, and the above theory is completely equivalent to the traditional Nakajima-Zwanzig approach. On the other hand, infinite dimensional systems, or even high-dimensional systems, do not allow to consider all possible sets of dynamical variables. In this case the set of relevant variables must be limited. 

Let us assume that the open system consists of one component or several weakly interacting components, labelled by $\alpha$.
The annihilation operator  $a_\alpha,$ the creation operator $a_\alpha^\dagger$ and the occupation numbers operator $n_\alpha=a^\dagger_\alpha a_\alpha$ can be chosen as relevant variables for such a system  \cite{VAL}. In other words, this set includes momentum or position and energy of the $\alpha$th component of the open system, which are directly accessible to experiments.  In the case of strongly interacting parts of an open system individual characteristics of the parts are not relevant and the set of observables has to be chosen as $a_\alpha^\dagger a_\beta,$ $a_\alpha^\dagger a_\beta^\dagger,$ and $a_\alpha a_\beta$ \cite{REPKE}. 
Thus, the relevant distribution \eqref{reldist} for an open system, can be written as follows
\begin{equation}
\rho_{\mathrm{rel}}=\frac{1}{Z_1Z_2}\exp[-\sum_m F_m P_m]\otimes \exp[-\beta H_B],
\end{equation}
where $H_B$ is the Hamiltonian of the thermostat, $P_m=\{a_i, a^\dagger_i a_i, a^\dagger_i\}\,\mathrm{or}\, \{a_i^\dagger a_j, a_i^\dagger a_j^\dagger,a_i a_j \}$ is the set of relevant variables, which characterise the $i$th subsystem of the open system, and $\Phi_i=\log Z_i$ are the Massieu-Planck functions of the subsystems.

Below we present examples to illustrate the suggested approach. 

\section{Damped harmonic oscillator} 
The simplest model of an open system is the damped harmonic oscillator. The rotating wave Hamiltonian of the system is 
\begin{displaymath}
H=\omega_0 a^\dagger a+\sum_j\omega_j b_j^\dagger b_j+\sum_jg_j(a b_j^\dagger+a^\dagger b_j),
\end{displaymath}
where $a$ and $a^\dagger$ are the boson annihilation and creation operators, $b_j$ and $b_j^\dagger$ are the annihilation and creation operators of $j$th oscillator in the bath, $\omega_0$ and $\omega_j$ are the frequencies of the oscillator and  $j$th oscillator in the bath and $g_j$ are the interaction strengths.The first step of the formalism is to choose the relevant distribution. The open system consists of only one component, but the dimension of the open system space is infinite. So, we choose as the relevant variables the free Hamiltonian of the bath $H_B=\sum_j\omega_j b_j^\dagger b_j,$ the free Hamiltonian of the oscillator $H_0=\omega_0 a^\dagger a$ and, also, the annihilation and creation operators $a, \, a^\dagger$. The relevant distribution can be written as
\begin{equation}
\rho_\mathrm{rel}\!\!=\!\!\frac{1}{Z_1Z_2}\!\exp[-\!F_1(t)a^{\!\dagger}\!\!-\!\!F_2(t) a^{\!\dagger}\! a\!\!-\!\!F_3(t)a]\exp[-\beta H_B].
\end{equation}
In the above equation $Z_1=\mathrm{Tr}\exp[-F_1(t)a^\dagger-F_2(t) a^\dagger a-F_3(t)a]=\exp[F_2(t)+F_1(t)F_3(t)/F_2(t)]/(\exp[F_2(t)]-1)$ and $Z_2=\mathrm{Tr}\exp[-\beta H_B],$ $\beta$ is the inverse thermodynamical temperature of the bath.

The self-consistency conditions \eqref{scond} can be easily solved and the result is
\begin{eqnarray}
F_1(t)=-F_2(t)\langle a\rangle^t, \,\,\, \,\,\, F_3(t)=-F_2(t)\langle a^\dagger\rangle^t, \label{Osc_SC} \\
F_2(t)=\ln\frac{\langle a\rangle^t\langle a^\dagger\rangle^t-\langle a^\dagger a\rangle^t-1}{\langle a\rangle^t\langle a^\dagger\rangle^t-\langle a^\dagger a\rangle^t}.\label{Osc_SC1}
\end{eqnarray}
Now an explicit formula for the entropy \eqref{entropy} in terms of the mean values of the relevant variables can be derived
\begin{eqnarray}\label{Osc_entropy}
S(t)&=&-(\langle a^\dagger a\rangle^t-\langle a\rangle^t\langle a^\dagger\rangle^t)\ln(\langle a^\dagger a\rangle^t-\langle a\rangle^t\langle a^\dagger\rangle^t)\\
&&\!\!\!\!\!\!\!\!\!\!\!\!\!\!\!\!\!\!\!\!\!\!+(1\!+\!\langle a^\dagger a\rangle^t\!-\!\langle a\rangle^t\langle a^\dagger\rangle^t)\ln(1+\langle a^\dagger a\rangle^t-\langle a\rangle^t\langle a^\dagger\rangle^t)+S_\mathrm{eq}\nonumber,
\end{eqnarray}
where, $S_\mathrm{eq}$ is the equilibrium entropy of the bath.
This fundamental result arises automatically as a part of the formalism. Such a simple connection of dynamics and thermodynamics is a significant feature of the considered method. In fact, Eq. ~\eqref{Osc_entropy} is a generalization of the informational entropy to the non-Markovian and non-equilibrium cases for the damped harmonic oscillator.

The knowledge of the thermodynamical entropy allows to determine other thermodynamical parameters. For instance, the non-equilibrium inverse temperature is $\beta(t)=\frac{\partial S}{\partial \langle H_S\rangle_t}$ by definition \cite{ZUB}. One can check that for  the system under consideration $\beta(t)=\frac{\partial S}{\partial \langle H_S\rangle_t}=F_2/\omega_0$.

The form of the above expression for the non-equilibrium entropy does not depend on details of the dynamical evolution. To apply this expression it is necessary to determine the non-equilibrium average of the relevant variables. The dynamics of the relevant observables is governed by Eq.~\eqref{local}. For the considered model, up to terms of the second order in the coupling constant, we obtain
\begin{eqnarray}
\frac{\partial \langle a\rangle^t}{\partial t}\!\!&=&\!\!-\langle a\rangle^{t}f^*(t) \label{Osc_trans}
,\\
\frac{\partial \langle a^\dagger a \rangle^t}{\partial t}\!\!\!&=&-\left[f(t)+f^*(t)\right] \langle a^\dagger\! a \rangle^{t}\nonumber\\
&+&\frac{1}{2}\left[f(t,\beta)+f^*(t,\beta)-f(t)-f^*(t)\right] \label{Osc_trans1}. 
\end{eqnarray} 
In the above equations we introduced the following correlation functions
\begin{eqnarray*}
f(t,\beta)=\int_0^t dt'\int_0^\infty d\omega J(\omega)\coth\left(\frac{\beta \omega}{2}\right)e^{i(\omega-\omega_0)t'}, \\
f(t)=f(t,\infty)=\int_0^t dt'\int_0^\infty d\omega J(\omega)e^{i(\omega-\omega_0)t'},
\end{eqnarray*}
where $J(\omega)$ is the spectral density of the bath.

Eqs.~\eqref{Osc_trans} and \eqref{Osc_trans1} seem identical to equations derived with the help of the traditional TCL master equation \cite{toqs} for the model. Nevertheless, the solution of Eqs.~\eqref{Osc_trans} and \eqref{Osc_trans1} completely determines the non-equilibrium dynamics to second order in the coupling constant as opposed to the traditional approach, where for the model one has to solve an infinite system of the dynamical equations. 
To clarify the difference let us determine the correlation function $\langle a^\dagger a a^\dagger a\rangle ^t=\mathrm{Tr}a^\dagger a a^\dagger a\rho_\mathrm{rel}=2{[\langle n\rangle^t]}^2+\langle n\rangle^t-[\langle a\rangle^t\langle a^\dagger\rangle^t]^2.$ Clearly, to find this correlation function from the traditional master equation one has to find a dynamical equation for the function and resolve it.

The solution of Eqs.~\eqref{Osc_trans}-\eqref{Osc_trans1} can be written as $\langle a\rangle^t=$ $\langle a\rangle^0\exp\left(-\int_0^tf^*(t')dt'\right)$ and $\langle a^\dagger a\rangle^t=$ $\langle a^\dagger a\rangle^0\exp\left(-\int_0^t(f^*(t')+f^*(t'))dt'\right)+$ $\frac{1}{2}\int_0^t\exp\left(-\int_s^t(f^*(t')+f^*(t'))dt'\right)K(s)ds,$where $K(t)$ is the free term in Eq.~\eqref{Osc_trans1}. Now, choosing the spectral density as $J(\omega)=\omega e^{-\omega/W},$ where $W$ is the cutoff frequency, and substituting the solution of the equations  \eqref{Osc_trans} and \eqref{Osc_trans1} in \eqref{Osc_entropy} we find the non-equilibrium entropy. Moreover, %, as it is known \cite{Repke}, 
the inverse temperature is %thermodynamically conjugate to the energy of the system, so for the damped harmonic oscillator we have 
$\beta(t)=F_2(t)/\omega_0.$ Thus, we described the thermodynamical properties of the open system in addition to the dynamical properties. The typical evolution of the thermodynamical entropy and the inverse temperature is shown in Fig.~1. The Markovian evolution corresponds to the parameters $f(\infty, \beta)$ and  $f(\infty)$ . The system relaxes from the initial state $\langle a^\dagger a\rangle^{0}=9$ and $\langle a\rangle^{0}=1.$ It is clearly seen that the thermodynamical parameters go to the equilibrium value, and the equilibrium temperature is equal to the temperature of the bath.  In the non-Markovian case the thermodynamical parameters reach equilibrium faster and oscillate around the equilibrium value.

\begin{figure}
\includegraphics[scale=0.7]{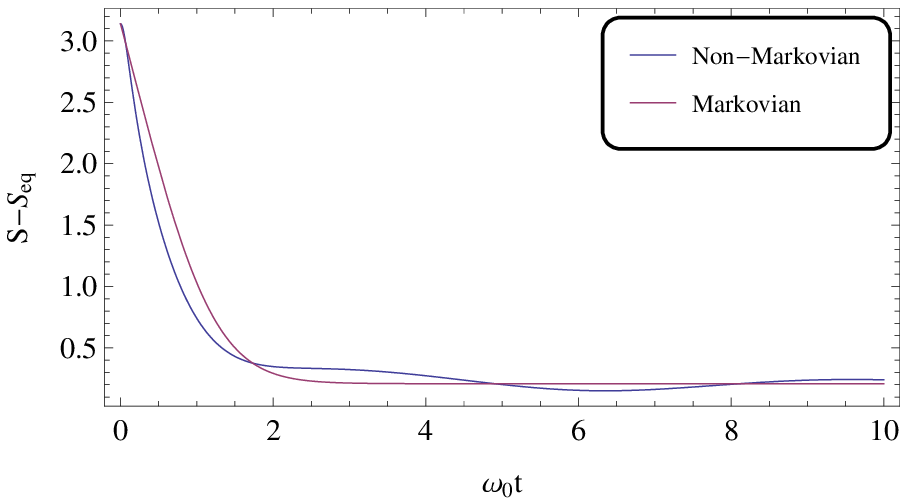}
\includegraphics[scale=0.7]{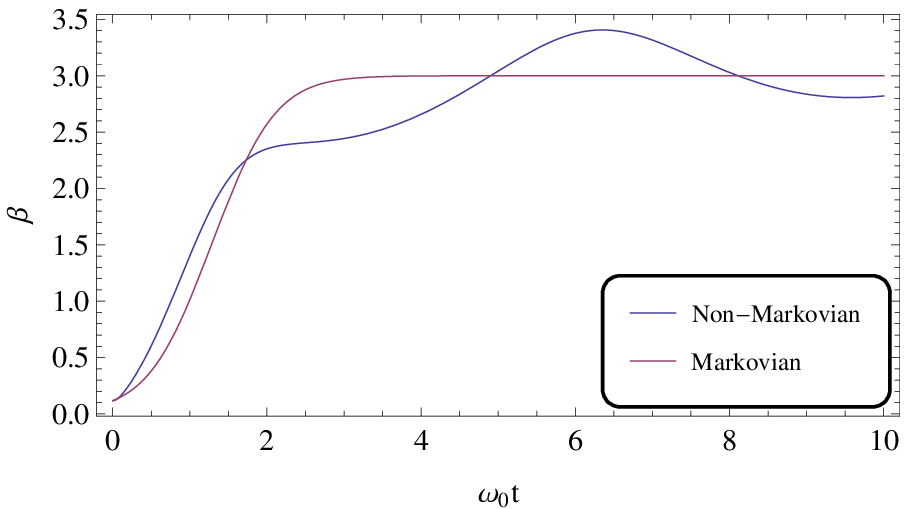}
\caption{Entropy (top) and inverse temperature (bottom) for the damped oscillator. Parameters in the system: $W=10,\, \beta_{\text{bath}}=3.$}
\end{figure}

\section{Driven two-level system} 
The second example which we consider is a damped driven two-level system. The Hamiltonian of the system is $H=\omega_0\sigma_z+(\Omega(t)\sigma_++\Omega^*(t)\sigma_-)+\sum_j\omega_j b_j^\dagger b_j+\sum_jg_j(\sigma_- b_j^\dagger+\sigma_+ b_j),$ where $\Omega(t)=\frac{\Omega}{2} e^{-i \omega_L t}$ is monochromatic classical field with frequency $\omega_L$ and Rabi frequency $\Omega,$  $\sigma_\pm=(\sigma_x\pm i\sigma_y),$ and the $\sigma_i$ are spin matrices. In this case the open system has a finite number of degrees of freedom. As relevant variables we can choose all possible observables. The relevant distribution has the form
\begin{equation}
\rho_\mathrm{rel}\!\!=\!\!\frac{1}{Z_1Z_2}\!\exp[-\!F_1\!(t)\sigma_+\!\!-\!\!F_2\!(t)\sigma_z\!\!-\!\!F_3\!(t)\sigma_-]\!\exp[\!-\!\beta H_B\!],\!\!
\end{equation}
where  %$\Phi_1=\ln 
$Z_1=\mathrm{Tr}\exp[-\!F_1\!(t)\sigma_+\!\!-\!\!F_2\!(t)\sigma_z\!\!-\!\!F_3\!(t)\sigma_-]=2\cosh(\frac{1}{2}\sqrt{4F_1(t)F_3(t)+F_2(t)^2})$ and  $Z_2=\mathrm{Tr}\exp[-\beta H_B].$ The solution of the self-consistency conditions \eqref{scond}  is
\begin{equation}\label{SCTL}
F_1(t)\!\!=\!\!-\langle\sigma_-\!\rangle^t\! R(t), \,F_2(t)\!\!=\!\!-2\langle\sigma_z\!\rangle^t \!R(t),\, F_3(t)\!\!=\!\!F_1^*(t),\!\!\!\!
\end{equation}
where we introduced the functions $R(t)=\mathrm{arctanh}(2X)/X,$ and $X=\sqrt{|\langle\sigma_-\rangle^t|^2+(\langle\sigma_z\rangle^t)^2}$. Thus, the non-equilibrium entropy \eqref{entropy} of the system has the form
\begin{equation}\label{entropytl}
S(t)=-2X^2R(t)+\ln 2-1/2\ln(1-4X^2)+S_\mathrm{eq}. 
\end{equation}

The expression \eqref{entropytl} for the thermodynamical entropy  is valid for the non-Markovian and the non-equilibrium cases. This result generalises the results obtained in \cite{TLENT} for Markovian dynamics. 

The dynamics of the two-level system is governed by the master equation of the form \eqref{local}. For the model under consideration the  equations up to the second order in the coupling constant are written as 
\begin{eqnarray}
\frac{\partial \langle\sigma_z\rangle^t}{\partial t}&=&i(\Omega\langle\sigma_-\rangle^t-\Omega\langle\sigma_+\rangle^t) \label{tlt1}\\
&-&\!\!\!\!\! \langle\sigma_z\rangle^{t}\left(f(t,\beta)+f^*(t,\beta)\right)-\!\frac{1}{2}\left(f(t)+f^*(t)\right),  \nonumber\\
\frac{\partial \langle\sigma_+\rangle^t}{\partial t}&=&-2i\Omega\langle\sigma_z\rangle^t-f(t,\beta) \langle\sigma_+\rangle^{t}. \label{tlt2}
%&\!\!\!\!\!\!\!\!\!\!\!\!\!\!\!\!\!\!\!\!+&\!\!\!\!\!\!\!\!\!\!\!\!\!\!\!\!f(t,\beta) \langle\sigma_+\rangle^{t} \nonumber
\end{eqnarray}
In the above equations the functions $f(t,\beta)$ and $f(t)$ are the same as for the previous model. The dynamical equations \eqref{tlt1}-\eqref{tlt2} are identical to the usual result derived with the help of the TCL master equation \cite{toqs}. For this model the two approaches give the same results.  This is due to the fact that we considered all possible dynamical variables of the open system as relevant.

The particular choice of the spectral density allows to solve the equations \eqref{tlt1}-\eqref{tlt2} numerically. The entropy of the open system is given by \eqref{entropytl} and the non-equilibrium inverse temperature is $\beta=F_2(t)/\omega_0,$ where $F_2(t)$ is defined by \eqref{SCTL}.
Thermodynamical entropy and inverse temperature, which correspond to spectral density $J(\omega)=\omega e^{-\omega/W},$ are shown in Fig.~2. The Markovian evolution corresponds to the functions $f(\infty)$ and $f(\infty,\beta),$ the system relaxes from the initial state $ \langle\sigma_z\rangle^0= \langle\sigma_+\rangle^0= \langle\sigma_-\rangle^0=0.$  It is clearly seen that the thermodynamic parameters relax to the same quasi-equilibrium state in both the Markovian and non-Markovian cases. At the same time, in the non-Markovian case $S(t)$ and $\beta(t)$ have oscillatory nature at sufficiently large times, while Markovian dynamics does not show such behaviour.  The quasi-equilibrium temperature of the open system is higher than the corresponding equilibrium temperature of the bath (we plotted the inverse temperature in Fig. 2). This effect can be easily understood if we remember that the system interacts with the external classical field. The  classical  field carries the energy to the system, in other words the system is continuously heated. This contribution of the external field is responsible for the difference between the equilibrium temperature of the bath and the quasi-equilibrium temperature of the open system.

\begin{figure}
\includegraphics[scale=0.7]{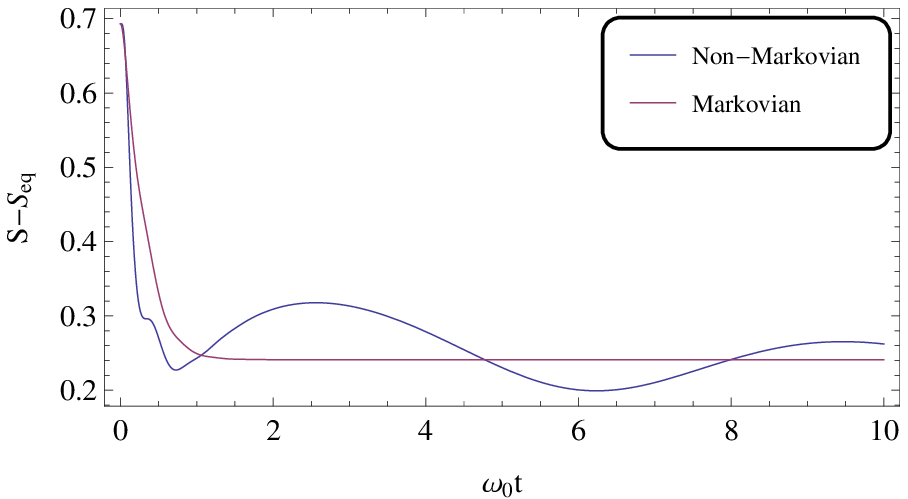}
\includegraphics[scale=0.7]{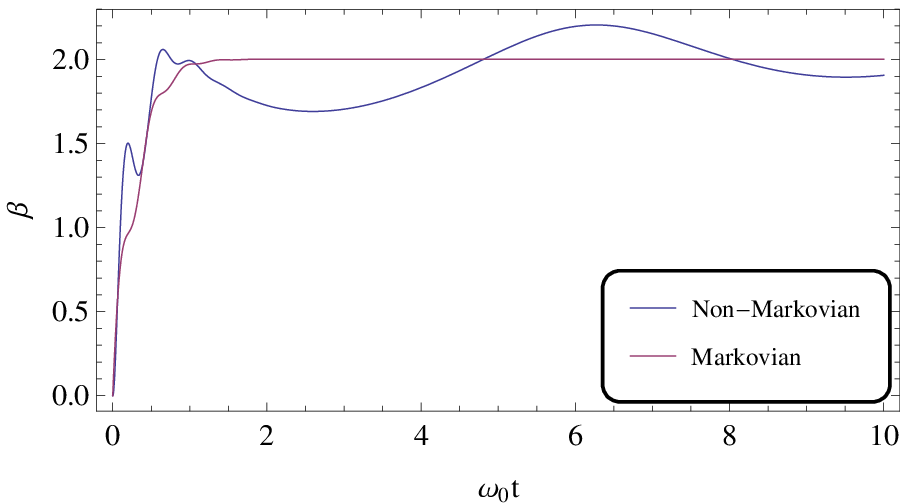}
\caption{Entropy (top) and inverse temperature (bottom) for the driven two-level system. Parameters in the system: $W=10,\, \beta_{\text{bath}}=3, \, \Omega=5.$}
\end{figure}

\section{Conclusion} 
In this article we have suggested to describe open quantum systems by unifying ideas of non-equilibrium thermodynamics and the time-convolutionless master equation. As  significant features of the approach we can indicate the following: (i) the method allows to study only necessary dynamical variables, in contrast to Nakajima-Zwanzig and related techniques, which either use a full set of the system variables or a diagonal subset. In this sense the suggested method is more flexible; (ii) the method includes a connection between dynamical and thermodynamical characteristic of an open system, given by the self-consistency conditions \eqref{scond}. The solution of the self-consistency condition \eqref{scond} is much easier than the direct calculation of thermodynamical parameters using the definition \eqref{entropy};
(iii) the main non-Markovian master equation \eqref{local} is local in time and can be efficiently studied with the help of usual methods of analysis; (iv) the approach can be easily modified to take into account a non-stationarity of a bath state.

\begin{acknowledgments}
This work is based upon research supported by the South African
Research Chair Initiative of the Department of Science and
Technology and National Research Foundation.
\end{acknowledgments}

\appendix

\section{Properties of the Kawasaki-Gunton projection operator.}
Let us firstly indicate the significant properties of the Kawasaki-Gunton projection operators (Eq.~(4))
\begin{equation}\label{kawas1}
\mathcal{P}\!(t)A\!\!=\!\!\rho_{\mathrm{rel}}\!(t)\mathrm{Tr}A\!+\!\!\!\sum_m\!\! \left\lbrace\mathrm{Tr}\!(AP_m)\!\!-\!\!(\mathrm{Tr}\!A)\!\langle\! P_m \rangle^t\!\right\rbrace\!\!\frac{\partial\rho_\mathrm{rel}(t)}{\partial \langle P_m \rangle^t},
\end{equation}
namely
\begin{eqnarray}
\mathcal{P}(t)\mathcal{P}(t')=\mathcal{P}(t),\\
\mathcal{P}(t)\rho(t)=\rho_\mathrm{rel}(t),\\
\mathcal{P}(t)\frac{\partial \rho(t)}{\partial t}=\frac{\partial \rho_\mathrm{rel}(t)}{\partial t}.
\end{eqnarray}
In the above expressions $\rho(t)$ is an arbitrary density operator. The proof of these expressions can be found in \cite{Repke}.

\section{The local in time master equation.}
Through action of the Kawasaki-Gunton projection operator on both side of the Liouville equation $\dot{\rho}=L\rho$ we derive the system 
\begin{eqnarray}
\mathcal{P}(t)\frac{\partial \rho(t)}{\partial t}=\mathcal{P}(t)L(t)\mathcal{P}(t)\rho(t)+\mathcal{P}(t)L(t)\mathcal{Q}(t)\rho(t),\label{1}\\
\mathcal{Q}(t)\frac{\partial \rho(t)}{\partial t}=\mathcal{Q}(t)L(t)\mathcal{P}(t)\rho(t)+\mathcal{Q}(t)L(t)\mathcal{Q}(t)\rho(t),\label{2}
\end{eqnarray}
where $\mathcal{Q}(t)=1-\mathcal{P}(t)$ is the additional operator. The formal solution of Eq.~ \eqref{2} can be written as
\begin{eqnarray}
\mathcal{Q}(t)\rho(t)&=&\mathcal{G}(t,t_0){Q}(t_0)\rho(t_0)\nonumber\\
&+&\int_{t_0}^t ds \mathcal{G}(t,s){Q}(s)L(s)\mathcal{P}(s)\rho(s),
\end{eqnarray}
where $\mathcal{G}(t,t')=\mathcal{T}_-\exp\left\lbrace\int\limits_{t'}^{t}\mathcal{Q}(\tau)L(\tau)d\tau\right\rbrace $ is the propagator. Substituting this result in \eqref{1} we obtain the generalized Kawasaki-Gunton master equation.

The  master equation local in time is derived by inversion of the solution of the initial Liouville equation $\rho(s)=G(t,s)(\mathcal{Q}(t)+\mathcal{P}(t))\rho(t)$ and substituting this expression for (B3). The explicit form of the superoperator $G(t,s)$ is given in the main text. After some algebra we derive
\begin{equation}
[1-\Sigma(t)]\mathcal{Q}(t)\rho(t)=\mathcal{G}(t,t_0)\mathcal{Q}(t_0)\rho(t_0)+\Sigma(t)\mathcal{P}(t)\rho(t).
\end{equation}
Multiplying both sides of the above equation by  $[1-\Sigma(t)]^{-1}$ and substituting the result for \eqref{1} we obtain the local in time master equation (Eq.~(5) in the main text).

Expanding the right-hand side of Eq.~(5) in the main text with respect to the small interaction constant up to the second order we obtain
\begin{widetext}
\begin{eqnarray}
\mathcal{K}(t)=\mathcal{P}(t)L(t)\mathcal{P}(t)+\int_{t_0}^t[\mathcal{P}(t)L(t)L(t_1)\mathcal{P}(t)-\mathcal{P}(t)L(t)\mathcal{P}(t_1)L(t_1)\mathcal{P}(t)]dt_1,\label{mat1}\\
\mathcal{I}(t)=\mathcal{P}(t)L(t)\mathcal{Q}(t_0)+\int_{t_0}^t[\mathcal{P}(t)L(t)L(t_1)\mathcal{Q}(t_0)-\mathcal{P}(t)L(t)\mathcal{P}(t_1)L(t_1)\mathcal{Q}(t_0)]dt_1\label{mat2}.
\end{eqnarray}
\end{widetext}

The expressions \eqref{mat1}-\eqref{mat2} is the generalised local in time master equation of the second order, which holds for any projection operator. Particularly, time-independent projection operators yield the well-known TCL master equation \cite{toqs}. The usual simplification for such equations is a special choice of initial conditions, namely, $\rho(t_0)=\mathcal{P}(t_0)\rho(t_0),$ and it leads to absence of the inhomogeneity in the master equation.
 This simplification was used in the main text.

\section{Generalized transport equations.}

To use the Kawasaki-Gunton projection operator technique it is convenient to multiply both sides of the master equation by some relevant variable $P_n$ and to take the trace, after that, using the linearity of the derivatives and trace and self-consistency conditions, the master equation can be re-written as a so-called generalized transport equation. Performing this operation for all relevant variables one obtains a system of generalized transport equations. 

As example let us consider the damped harmonic oscillator, which is the first model in the main text. The Hamiltonian in the interaction picture is $H_I(t)=\sum_j g_j(ab^\dagger_j e^{-i(\omega_0-\omega_j)t}+a^\dagger b_j e^{i(\omega_0-\omega_j)t})=(a B^\dagger(t) e^{-i\omega_0t}+a^\dagger B(t) e^{i\omega_0t}),$  where $B(t)=\sum_j g_j b_j e^{-i\omega_j t}.$ It is clear that $\mathcal{P}(t)[H_I(t),\mathcal{P}(t)\rho(t)]=0.$ So, the second order master equation for the model is 
\begin{equation}\label{eq}
\dot{\rho}_\mathrm{rel}=-\int_{t_0}^t\mathcal{P}(t)[H(t),[H(t_1),\rho_\mathrm{rel}(t)]]dt_1.
\end{equation}
Using the cyclic permutation under the trace sign it is easy to show that $\mathrm{Tr}L(t)A=0$ for arbitrary operator $A$. Thus, in the above equation the Kawasaki-Gunton projection operator can be replaced by the Robertson projection operator, which acts as \begin{equation}\label{robert}
\mathcal{P}_R\!(t)A\!\!=\sum_m\!\! \mathrm{Tr}\!(AP_m)\frac{\partial\rho_\mathrm{rel}(t)}{\partial \langle P_m \rangle^t}.
\end{equation}
Multiplication of both sides of the master equation \eqref{eq} by a relevant variable $P_m$ and taking the trace one obtains
\begin{equation}\label{eq2}
\dot{\langle P_m\rangle^t}=-\int_{t_0}^t\mathrm{Tr}\left\lbrace [H(t),[H(t_1),\rho_\mathrm{rel}(t)]]dt_1 P_m\right\rbrace,
\end{equation}
where we used $\mathrm{Tr}\frac{\partial\rho_\mathrm{rel}(t)}{\partial \langle P_n \rangle^t}P_m=\frac{\partial}{\partial \langle P_n \rangle^t}\mathrm{Tr}\rho_\mathrm{rel}P_m=\frac{\partial\langle P_m \rangle^t}{\partial \langle P_n \rangle^t}=\delta_{mn}.$ Substituting for $P_m$ specific relevant variables, namely $a,$ $a^\dagger,$ and $a^\dagger a,$ leads to the generalized transport equations (Eqs.~(13) and (14)). Notice, that we used the following expression for the bath correlation functions 
\begin{widetext}
$$\mathrm{Tr}B(t) B^\dagger(t_1) e^{-\beta H_B}/Z_2=\int_0^\infty d\omega e^{-i\omega(t-t_1)}J(\omega)[\coth(\omega \beta/2)+1]/2$$ and $$\mathrm{Tr}B^\dagger(t) B(t_1) e^{-\beta H_B}/Z_2=\int_0^\infty d\omega e^{i\omega(t-t_1)} J(\omega)[\coth(\omega \beta/2)-1]/2. $$
\end{widetext}

To study the second model we transform the initial Hamiltonian to the interaction picture with respect to the Hamiltonian $$H_0=\omega_0\sigma_z+(\Omega(t)\sigma_++\Omega^*(t)\sigma_-)+\sum_j\omega_j b_j^\dagger b_j.$$ 
The corresponding evolution operator is 
\begin{widetext}
$$U(t,t')\!=\!\mathcal{T}_-\exp\left\lbrace-i\int\limits_{t'}^{t}H_0 d\tau\right\rbrace\!=\!(- d^*(t,t') \sigma_+\!+d(t,t')\sigma_-\!+(c(t,t')-c^*(t,t'))\sigma_z+(c(t,t')+c^*(t,t'))/2)\otimes e^{-i \sum_j\omega_j b_j^\dagger b_j(t-t') },$$
 where
 \begin{eqnarray*}
c(t,t')\!=\!e^{-\frac{i}{2}(t-t')\omega_L}\!\!\!\left(\!\!\cos[\frac{k}{2}(t-t')]\!+\!\frac{i (\omega_L-\omega_0)}{k}\sin[\frac{k}{2}(t-t')] \!\!\right)\!\!,\nonumber\\
d(t,t')\!=\!\frac{\Omega}{ik}e^{\frac{i}{2}(t+t')\omega_L}\sin[\frac{k}{2}(t-t')], \,
k\!=\!\!\sqrt{\Omega^2+(\omega_L-\omega_0)^2}.\nonumber
\end{eqnarray*}
The transformed Hamiltonian $H_I(t)=U^\dagger(t,0)(\sigma_- B^\dagger+\sigma_+ B )U(t,0)$ is defined as 
\begin{eqnarray}
H_I(t)&=&2(d(t,0)c^*(t,0) B(t)+d^*(t,0)c(t,0) B^\dagger(t))\sigma_z \label{ham2}\\
&+&((c^*(t,0))^2B(t)-(d^*(t,0))^2B^\dagger(t))\sigma_+-(d(t,0)^2B(t)-c^2(t,0)B^\dagger(t)(t))\sigma_-.\nonumber
\end{eqnarray}
\end{widetext}
The generalized transport Eqs.~(18) and (19) are the particular case of Eq.~\eqref{eq2} with the Hamiltonian \eqref{ham2} in the case of the exact resonance $\omega_0=\omega_L$ and not very strong intensity of the classical field $\Omega\ll\omega_0.$

\section{Thermodynamical properties of a system.}
To describe the thermodynamical properties of an open system one has to solve the self-consistency conditions $\langle\! P_m \rangle^t=\mathrm{Tr} (P_m\rho_\mathrm{rel}).$ Formally, the non-equilibrium averages $\langle\! P_m \rangle^t$ are known and one has to resolve the conditions with respect to $F_m(t).$ 

The simplest way to derive an explicit form of the self-consistency conditions is to  calculate the Massieu-Planck function $\Phi(t).$ In this case the required relations are given by $\langle P_m\rangle^t=-\frac{\partial \Phi}{\partial F_m }$ (Eqs.~(7) of the main text). 

For the damped harmonic oscillator we have $\Phi(t)=\ln Z_1=\ln\exp[F_2(t)+F_1(t)F_3(t)/F_2(t)]/(\exp[F_2(t)]-1),$ which can be found in the main text. Differentiation of the Massieu-Planck function with respect to $F_m(t)$ gives 
\begin{eqnarray}
\!\!\!\!\!\!\!\!\langle a^\dagger\rangle^t&=&-\frac{\partial \Phi}{\partial F_1 }=-F_3(t)/F_2(t),\\
\!\!\!\!\!\!\!\!\langle a^\dagger a\rangle^t\!\!\!\!&=&-\frac{\partial \Phi}{\partial F_2 }=1/(e^{F_2(t)}-1)\!+\!F_1(t)F_3(t)/F_2^2(t)\!, \\
\!\!\!\!\!\!\!\!\langle a\rangle^t&=&-\frac{\partial \Phi}{\partial F_3 }=-F_1(t)/F_2(t).
\end{eqnarray}
The solution of the system above is presented in the main text (Eqs.~ (10) and (11) of the main text).

For the driven two-level system we have $\Phi(t)=\ln Z_1=\ln(2 \cosh (\sqrt{4F_1(t)F_3(t)+F_2(t)^2}/2))$. Differentiation of the Massieu-Planck function with respect to $F_m(t)$ gives 
\begin{widetext}
\begin{eqnarray}
\langle \sigma_+\rangle^t&=&-\frac{\partial \Phi}{\partial F_1 }=-F_3(t)\tanh(\sqrt{4F_1(t)F_3(t)+F_2(t)^2}/2)/\sqrt{4F_1(t)F_3(t)+F_2(t)^2},\\
\langle \sigma_z\rangle^t&=&-\frac{\partial \Phi}{\partial F_2 }=-F_2(t)\tanh(\sqrt{4F_1(t)F_3(t)+F_2(t)^2}/2)/(2\sqrt{4F_1(t)F_3(t)+F_2(t)^2}), \\
\langle \sigma_-\rangle^t&=&-\frac{\partial \Phi}{\partial F_3 }=-F_1(t)\tanh(\sqrt{4F_1(t)F_3(t)+F_2(t)^2}/2)/\sqrt{4F_1(t)F_3(t)+F_2(t)^2}.
\end{eqnarray}
\end{widetext}
The solution of the system above is presented in Eq.~(16).

\section{Spectral density of the bath.}
A remarkable property of the spectral density of the bath $J(\omega)=\omega e^{-\omega/W}$, is the existence of an analytical form  of an inverse Fourier transformation, even for non-zero temperature of the bath. Namely,
\begin{widetext}
\begin{eqnarray}
\int_0^\infty d\omega J(\omega)\coth\left(\frac{\beta \omega}{2}\right)e^{i(\omega-\omega_0)t}=e^{-i t\omega_0} \left(\frac{W^2}{(t W+i)^2}+\frac{2 \psi '\left(\frac{1-i t W}{W \beta }\right)}{\beta ^2}\right), \,\,\,
\int_0^\infty d\omega J(\omega)e^{i(\omega-\omega_0)t}=\frac{e^{-i t \omega_0}}{\left(\frac{1}{W}-i t\right)^2},
\end{eqnarray}
\end{widetext}
where $\psi '\left(x\right)=\frac{d\psi\left(x\right)}{dx}$ and $\psi\left(x\right)$ is the Euler digamma function.

\end{document}